\documentstyle[aps,psfig]{revtex}

\newcommand{\be}{\begin{equation}}
\newcommand{\ee}{\end{equation}}
\newcommand{\ba}{\begin{eqnarray}}
\newcommand{\ea}{\end{eqnarray}}

\begin{document}
\title{{\bf Two-dimensional Bose-Einstein Condensation in Cuprate Superconductors}}
\author{M. Casas$^{a}$, M. de Llano$^{b}$, A. Puente$^{a}$, A. Rigo$^{a}$ and M.A.
Sol\'{\i}s$^{c}$}
\address{
$^{a}$Departament de F\'{\i}sica, Universitat de les Illes Balears,\\
07071 Palma de Mallorca, Spain \\
$^{b}$Instituto de Investigaciones en Materiales, Universidad Nacional\\
Aut\'{o}noma de M\'{e}xico,\\
Apdo. Postal 70-360, 04510 M\'{e}xico, DF, Mexico \\
$^{c}$Instituto de F\'{\i}sica, Universidad Nacional Aut\'{o}noma de\\
M\'{e}xico, 01000 M\'{e}xico, DF, Mexico.
}
\maketitle

\begin{abstract}
Transition temperatures $T_{c}$ calculated using the BCS model
electron-phonon interaction\ without any adjustable parameters agree with
empirical values for quasi-2D cuprate superconductors. \ They follow from a
two-dimensional gas of temperature-dependent Cooper pairs in chemical and
thermal equilibrium with unpaired fermions in a boson-fermion (BF)
statistical model as the Bose-Einstein condensation (BEC) singularity
temperature is approached from above. The {\it linear} (as opposed to
quadratic) boson dispersion relation due to the Fermi sea yields
substantially higher $T_{c}$'s with the BF model than with BCS or pure-boson
BEC theories.

\noindent

{\bf PACS \#} 03.75Fi; 05.30.-d; 05.30.Fk; 05.30.Jp \newline
\end{abstract}



We provide support to a widespread conjecture (or ``paradigm'') that
superconductivity in general is a Bose-Einstein condensation (BEC) of the
charged Cooper pairs \cite{Coo} (CPs) observed in magnetic flux quantization
experiments in classical \cite{classical,classical2} as well as cuprate \cite
{cuprates} superconductors. \ The same general conjecture is also often made
regarding the superfluidity of liquid helium-3 in terms of CPs consisting of
neutral-atom $^{3}$He\ fermions. \ BEC as a {\it statistical }(as opposed to%
{\it \ }a{\it \ dynamical}) mechanism\ of superconductivity has been
entertained, among others,\ by Anderson \cite{AndersonBEC}, by T.D. Lee \cite
{Lee1} and by Mott \cite{MottBEC}\ and their co-workers.\ But BEC normally
occurs only for dimensions $d>2$ while some modern superconductors are
quasi-2D or even quasi-1D materials. We show, however, that CPs can undergo
BEC for {\it all} $d>1$. We further obtain reasonable critical temperatures $%
T_{c}$\ without any adjustable parameters, thus bolstering the above
mentioned conjecture even before building in full many-body
self-consistency. As in BCS theory, fluctuations have also been neglected. \
More detailed, sophisticated treatments actually link \cite
{Lee2,Rann1,Rann2,Tolma,Rann3}.{\Huge \ }BEC (characterized by a {\it bosonic%
} {\it condensate fraction}) with the BCS theory (characterized by a {\it %
fermionic gap}), but report no attempts to calculate specific $T_{c}$'s
without adjustable parameters to compare with experiment.

A BEC picture of superconductivity is consistent with the recent discovery
of the ``pseudogap'' in the electronic density of states \cite
{ps1,pseudogap,ps2,ps3,ps3a,psb,ps4} above $T_{c}$ in certain cuprates, at
least with one of its major interpretations as ``pre-formed CPs'' without
long-range coherence or condensation, while in BCS theory CP formation and
condensation occur simultaneously below the {\it same} $T_{c}$. We here
submit that a natural candidate for such pre-formed CPs are the
nonzero-center-of-mass CPs usually neglected in BCS theory.\

To fix the dynamics take a 2D system of $N$ fermions of mass $m$ confined in
a square of area $L^{2}$ interacting pairwise via the BCS model
electron-phonon interaction $V_{{\bf k},{\bf k}^{\prime }}=-V$, with $V>0$,
whenever $\mu (T)-\hbar \omega _{D}<\epsilon _{k_{1}}$ $(\equiv \hbar
^{2}k_{1}^{2}/2m),\ \epsilon _{k_{2}}<\mu (T)+\hbar \omega _{D}$, and zero
otherwise, where ${\bf k}\equiv {\frac{1}{2}}({\bf k}_{1}-{\bf k}_{2})$ is
the relative wavevector of the two particles; $\mu (T)$ the ideal Fermi gas
(IFG) chemical potential, which at $T=0$ becomes the Fermi energy $%
E_{F}\equiv \hbar ^{2}k_{F}^{2}/2m$ with $k_{F}$ the Fermi wavenumber; and $%
\omega _{D}$ the Debye frequency. Striking direct evidence for significant
electron-phonon coupling in high-temperature cuprate superconductors from
angle-resolved photoemission spectroscopy (ARPES) experiments has recently
been reported \cite{Shen}.

If $\hbar {\bf K}=\hbar ({\bf k}_{1}+{\bf k}_{2})$ is the center-of-mass
momentum (CMM) of a CP, let $E_{K}$ be its {\it total} energy (besides the
CP rest-mass energy). The {\it original} CP \cite{Coo} eigenvalue\ equation
is then 
\begin{equation}
1=V{\sum_{{\bf k}}}^{^{\prime }}\frac{\theta (k_{1}-k_{F})\,\,\theta
(k_{2}-k_{F})}{2\epsilon _{k}+{\hbar ^{2}}K{^{2}/}4m-E_{K}},
\label{eq:cooper}
\end{equation}
where $\theta (x)$ is the Heaviside unit step function, and the prime on the
summation sign denotes the conditions $k_{1,2}\equiv |{\bf K/}2\pm {\bf k}%
|<(k_{F}^{2}+k_{D}^{2})^{1/2}\;$ensuring that the pair of particles {\it %
above} the Fermi ``surface'' cease interacting beyond the annulus of energy
thickness $2\hbar \omega _{D}$ $\equiv \hbar ^{2}k_{D}^{2}/m$, thereby
restricting the summation over ${\bf k}$ for a given fixed ${\bf K}$. CPs
obey BE statistics since there is an indefinitely large number of $\ {\bf k}$%
\ values in (\ref{eq:cooper}) for a given value of $K$. \ Setting $%
E_{K}\equiv 2E_{F}-\Delta _{K}$, a pair is {\it bound} if $\Delta _{K}>0$,
so that (\ref{eq:cooper}) becomes an equation for the (positive) pair
binding energy $\Delta _{K}$. Our $\Delta _{K}$ and $\Delta _{0}$ follow
Cooper's notation and should {\it not} be confused with the BCS energy gap $%
\Delta (T)$ at $T=0$. \ Let $g(\epsilon )$ be the electronic
density-of-states (for each spin) in the normal (i.e., interactionless) $N$%
-fermion state; in 2D it is constant, $g(\epsilon )=L^{2}m/2\pi \hbar
^{2}\equiv g$. The Cooper equation (\ref{eq:cooper}) for the unknown
quantity $\Delta _{K}$ can be analyzed beyond the usual zero-CMM, $K=0$,
case. For $K=0$ it becomes a single elementary integral, with the familiar 
\cite{Coo} solution $\Delta _{0}=2\hbar \omega _{D}/(e^{2/\lambda }-1)$
valid for {\it all} coupling $\lambda \equiv gV\geq 0$. For small $\lambda $
one gets \cite{PhysC} 
\begin{equation}
{\Delta _{K}}\mathrel{\mathop{\longrightarrow}\limits_{K \rightarrow 0}}%
\Delta _{0}-{\frac{2}{\pi }}\hbar v_{F}K+O(K^{2})  \label{eq:linear}
\end{equation}
where $v_{F}$ $\equiv \sqrt{2E_{F}/m}$ is the Fermi velocity. \ This {\it %
linear dispersion relation} is the 2D analog of the 3D result stated by
Schrieffer as far back as 1964 in Ref. \cite{Schrieffer}, p. 33 (see also
Ref. \cite{FW}, p. 336).\ Although some treatments (e.g., Ref. \cite{Janko})
of CPs more sophisticated than the original Cooper picture\ (\ref{eq:cooper}%
) numerically yield {\it resonant }pairs with a leading quadratic
dispersion, {\it linearly-dispersive resonances }appear analytically from a
Bethe-Salpeter equation many-body approach \cite{Fortes} to CPs in
3D---provided it is based on the BCS (where holes are treated on an equal
footing with particles), not the IFG, ground state.\ In 2D, see also Refs. 
\cite{Traven,Adh}.\ It is commonly confused with the also
linearly-dispersive sound phonons of the collective excitation sometimes
referred to as the Anderson-Bogoliubov-Higgs mode (which for zero coupling
reduces \cite{Belkhir} to the IFG result $\hbar v_{F}K/\sqrt{d}$). The IFG
sound speed $c=v_{F}/\sqrt{d}$ follows trivially from the zero-temperature
IFG pressure $P=n^{2}[d(E/N)/dn]=2nE_{F}/(d+2)$ via the familiar
thermodynamic relation $dP/dn=mc^{2}$, where $E$ is the ground-state energy
and $n\equiv N/L^{d}=k_{F}^{d}/2^{d-2}\pi ^{d/2}d\;\Gamma (d/2)$ is the
fermion-number density. But the simple result (\ref{eq:linear}) in fact
refers to actual ``moving'' (or ``excited'') CPs {\it in the Fermi sea,}
which clearly ``break up'' for $K>K_{0}$ as defined by $\Delta
_{K_{0}}\equiv 0$. {\it Both} kinds of {\it distinct} soundwave-like
solutions---moving CPs and ABH phonons---appear side by side in the
many-body, ladder-summation scheme of Ref. \cite{Fortes}.

For $N_{B}$ bosons of mass $m_{B}$ and energy $\varepsilon _{K}=C_{s}\,K^{s}$
with $s$\ $>0$ and $C_{s}$ a constant, a BEC temperature singularity occurs
at $T_{c}\neq 0$ for any dimension \cite{Gunton,Ziff} $d>s$ in the number
equation $N_{B}=\sum_{{\bf K}}[e^{(\varepsilon _{K}-\mu _{B})/k_{B}T}-1]^{-1}
$ at vanishing bosonic chemical potential $\mu _{B}\leq 0$ when the number
of ${\bf K}=0$ bosons just ceases to be negligible upon cooling. \ It is
given \cite{cas} by 
\begin{equation}
T_{c}=\frac{C_{s}}{k_{B}}\left[ \frac{s\,\Gamma (d/2)\,(2\pi )^{d}n_{B}}{%
2\pi ^{d/2}\,\Gamma (d/s)g_{d/s}(1)}\right] ^{s/d}  \label{Tc1}
\end{equation}
with $n_{B}\equiv $ $N_{B}/L^{d}$, and $g_{\sigma }(z)$ the usual Bose
integrals expandable as infinite series which are $\zeta (\sigma )$, the
Riemann zeta function of order $\sigma $, for $\sigma >1$ but diverge for $%
\sigma \leq 1$. Thus $T_{c}=0$ for all $d\leq s$. \ For $s=2$ and $d=3$ one
has $\zeta (3/2)\simeq 2.612$, and since $C_{2}\equiv $ $\hbar ^{2}/2m_{B}$ (%
\ref{Tc1}) then reduces to the familiar formula $T_{c}\simeq 3.31\hbar
^{2}n_{B}^{2/3}/m_{B}k_{B}$ of ``ordinary'' BEC. But for bosons with
(positive) excitation energy $\varepsilon _{K}\equiv \Delta _{0}-\Delta _{K}$
given approximately by the linear term in (\ref{eq:linear}) for all $K$,
meaning that $s=1$ and $C_{1}\equiv a(d)\hbar v_{F\text{ }}$ with $%
a(d)=2/\pi $ and $1/2$ for $d=2$ and $3$, respectively, the critical
temperature $T_{c}$ is {\it nonzero} {\it for all} $d$ $>{1}$--- {\it %
precisely} the dimensionality range of all known superconductors down to the
quasi-1D organo-metallic (Bechgaard) salts \cite{salts,salts2,salts3}.

The number of bosons in the boson-fermion mixture to be analyzed turns out
to be both coupling- and temperature-dependent and it is {\it in conserving
the fermion number} that a BEC-like singularity arises. As is the case for
the pure boson gas, a linear rather than a quadratic dispersion relation is
needed to obtain BEC in 2D. This emerges in a statistical model for the
ideal binary {\it mixture} of bosons (the CPs) and unpaired (both pairable
and unpairable) fermions in chemical equilibrium \cite{Schafroth,Schafroth2}
for which thermal pair-breaking into unpaired pairable fermions is
explicitly allowed \cite{PhysA}. Assuming the BCS model interaction the
total number of fermions in 2D at any $T$ is $N=L^{2}k_{F}^{2}/2\pi
=N_{1}+N_{2}$, where $N_{1}$ is the number of unpairable (i.e.,
non-interacting) fermions while $N_{2}$ is the number of pairable (i.e.,
active) ones. The unpairable fermions obey the usual Fermi-Dirac
distribution with the IFG chemical potential $\mu $ but the $N_{2}$ pairable
ones are simply those in the interaction shell of energy width $2\hbar
\omega _{D}$ so that, if $\ \beta \equiv (k_{B}T)^{-1}$, 
\begin{equation}
N_{2}=2\int_{\mu -\hbar \omega _{D}}^{\mu +\hbar \omega _{D}}\;d\epsilon 
\frac{g(\epsilon )}{e^{\beta (\epsilon -\mu )}+1}=2g\hbar \omega _{D},
\label{n214}
\end{equation}
which is independent of $T$. At fixed interfermionic coupling and
temperature these\ $N_{2}$\ fermions form an ideal mixture of pairable but
unpaired fermions plus CPs created near the single-fermion energy $\mu (T)$,
with binding energy $\Delta _{K}(T)$ $\geq 0$ and total energy 
\begin{equation}
E_{K}(T)\equiv 2\mu (T)-\Delta _{K}(T).  \label{bosonenergy}
\end{equation}
This generalizes the $T=0$ equation $E_{K}\equiv 2E_{F}-\Delta _{K}$
introduced before.

The Helmholtz free energy $F=E-TS$, where $E$ is the internal energy and $S$
the entropy, of this binary {\it ``composite
boson/pairable-but-unpaired-fermion system''} at temperatures $T\leq T_{c}$
is then readily constructed \cite{PhysA} in terms of: \ a) the average
number of unpaired but pairable fermions with fixed energy; b) $N_{B,K}(T)$,
the number of CPs with nonzero-CMM, $0<K\leq K_{0\text{,}}$ with the
CP-breakup value $K_{0}$ defined \cite{PhysC} by $\Delta _{K_{0}}\equiv 0$;
and c) $N_{B,0}(T)$, the number of CPs with zero CMM at temperature $T$. The
free energy $F_{2}$ of just the $N_{2}$ {\it pairable} fermions is to be
minimized subject to the constraint that $N_{2}$ is conserved, i.e., one
seeks the minimum of $F_{2}-\mu _{2}N_{2}$ with respect to (a), (b) and (c)
just mentioned. The total number of pairable but unpaired fermions $%
N_{20}(T) $ is then

\begin{equation}
N_{20}(T)=2g\int_{\mu -\hbar \omega _{D}}^{\mu +\hbar \omega _{D}}d\epsilon
\,\,{\frac{1}{e^{\beta (\epsilon -\mu _{2})}+1}} = {\frac{2g}{\beta }}\ln \left[ {\frac{1+e^{-\beta (\mu -\mu
_{2}-\hbar \omega _{D})}}{1+e^{-\beta (\mu -\mu _{2}+\hbar \omega _{D})}}}%
\right].  
\label{N20(T)F}
\end{equation}
The relevant {\it number equation} for the
pairable fermions is thus 
\begin{equation}
N_{2}=N_{20}(T)+2[N_{B,0}(T)+\sum_{K>0}^{K_{0}}N_{B,K}(T)]\equiv
N_{20}(T)+2N_{B}(T),  \label{N2}
\end{equation}
where $\sum_{K>0}^{K_{0}}N_{B,K}(T)=\sum_{K>0}^{K_{0}}[e^{\beta
\{E_{K}(T)-2\mu _{2}\}}-1]^{-1}$ is the {\it total} number of ``excited''
CPs (namely with CMM values $0<K<K_{0}$). One can rewrite $E_{K}(T)-2\mu
_{2} $ here as $\varepsilon _{K}{(T)}-\mu _{B}{(T)}$, with $\varepsilon _{K}{%
(T)}\equiv \Delta _{0}{(T)}-\Delta _{K}{(T)}\geq 0$ a (nonnegative)
excitation energy as suggested by (\ref{eq:linear}). \ Hole-hole and
particle-particle CPs can be shown to have the same excitation energy $%
\varepsilon _{K}{(}T{)}$. The remaining unknown $\mu _{B}{(T)}$ is then 
\begin{equation}
\mu _{B}(T)=2[\mu _{2}(T)-\mu (T)]+\Delta _{0}(T)=0  \label{N20(T)}
\end{equation}
for $0\leq T\leq T_{c}$\ since $N_{B,0}(T)$ is negligible for all $T>T_{c}$.
\ This is precisely the BEC condition for a {\it pure }boson gas, although
one now has a binary boson-fermion {\it mixture}.\ \

To determine $N_{B}(T)$ from (\ref{N20(T)F}) and (\ref{N2}) we use (\ref{N20(T)})\ and see that 

\begin{equation}
N_{20}(T)={\frac{2g}{\beta }}\ln \left[ {\frac{{1+e^{-\beta \{\Delta
_{0}(T)/2-\hbar \omega _{D}\}}}}{{1+e^{-\beta \{\Delta _{0}(T)/2+\hbar
\omega _{D}\}}}}}\right]  \label{n20t26}
\end{equation}
for $0\leq T\leq T_{c}$. Thus $2N_{B}(T)/N_{2}\equiv 1-N_{20}(T)/N_{2}$ is
obtainable for $0\leq T\leq T_{c}$ from (\ref{n20t26}) if $\Delta _{0}(T)$
were known. For this, $\theta (k_{1}-k_{F})\equiv \theta (\epsilon
_{k_{1}}-E_{F})$ 
in (\ref{eq:cooper}) becomes $1-n(\xi _{k_{1}})$ where $n(\xi
_{k_{1}})\equiv (e^{\beta \xi _{k_{1}}}+1)^{-1}$ with $\xi _{k_{1}}\equiv
\epsilon _{k_{1}}-\mu (T)$, the IFG chemical potential $\mu (T)$ in 2D being
given exactly by $\mu (T)=\beta ^{-1}\ln (e^{\beta E_{F}}-1)%
\mathrel{\mathop
{\longrightarrow}}E_{F}$ as $T\rightarrow 0$. Similar arguments hold for $%
\theta (k_{2}-k_{F})$. Since for $K=0$, $k_{1}=k_{2}$ which implies that $%
\xi _{k_{1}}=\xi _{k_{2}}$, (\ref{eq:cooper}) then provides a simple
generalization to finite-$T$ of the $K=0$ CP equation, namely 
\begin{equation}
1=\lambda \int_{0}^{\hbar \omega _{D}}d\xi (e^{-\beta \xi }+1)^{-2}[2\xi
+\Delta _{0}(T)]^{-1}.  \label{e13}
\end{equation}
Its numerical solution shows $\Delta _{0}(T)$ to decrease monotonically with 
$T$ for fixed $\lambda $ and $\hbar \omega _{D}$, and zero only at infinite $%
T$. \ (This infinite ``de-pairing'' temperature is obviously spurious as the
BCS model interaction loses meaning when $\mu (T)$ turns negative at large $%
T $.) \ Thus also $2N_{B}(T)/N_{2}$ decreases with $T$; it is plotted in
Fig. 1 as $2n_{B}(T)/n_{2},$ since $n_{B}(T)\equiv N_{B}(T)/L^{2}$ and $%
n_{2}\equiv N_{2}/L^{2}$.

Using (\ref{n214}) for $N_{2}$ the {\it fractional number of pairable
fermions that are actually paired} at $T=0$ becomes simply 
\begin{equation}
2N_{B}(0)/N_{2}=\Delta _{0}/2\hbar \omega _{D}=(e^{2/\lambda }-1)^{-1} %
\mathrel{\mathop{\longrightarrow} \limits_{\lambda \rightarrow 0}}
e^{-2/\lambda }  \label{NB/N2}
\end{equation}
for $\lambda \leq 2/\ln 2\simeq 2.89$, and $=1$ (all pairable fermions
paired into bosons) for $\ \lambda \geq 2.89$. This fraction is plotted
against coupling $\lambda $ in Fig. 1, and contrasts sharply with the
``heuristic model'' of Ref. \cite{cas}, Eq. (16), where $2N_{B}(0)/N_{2}$ $%
\equiv 1$ for {\it all }coupling. It is now more in line with BCS
theory---which is {\it not} \cite{Bardeen}\ a BEC theory---where, in any $d$%
, a coupling-dependent fraction is estimated (Ref. \cite{Blatt} p. 128) to
be $(\Delta /\hbar \omega _{D})^{2}\equiv (\sinh 1/\lambda )^{-2} %
\mathrel{\mathop{\longrightarrow}}4e^{-2/\lambda }$ as $\lambda \rightarrow
0 $. \ Here $\Delta $ (again, not to be confused with the CP binding energy $%
\Delta _{0}$) is the $T=0$ BCS energy gap for the same BCS model interaction
used in this Letter. It is graphed as the thin curve in Fig. 1 and is seen
to be much larger than (\ref{NB/N2}) for fixed $\lambda $.

\begin{figure}[tbh]
\begin{minipage}[b]{2.8in}
\begin{center}
\psfig{file=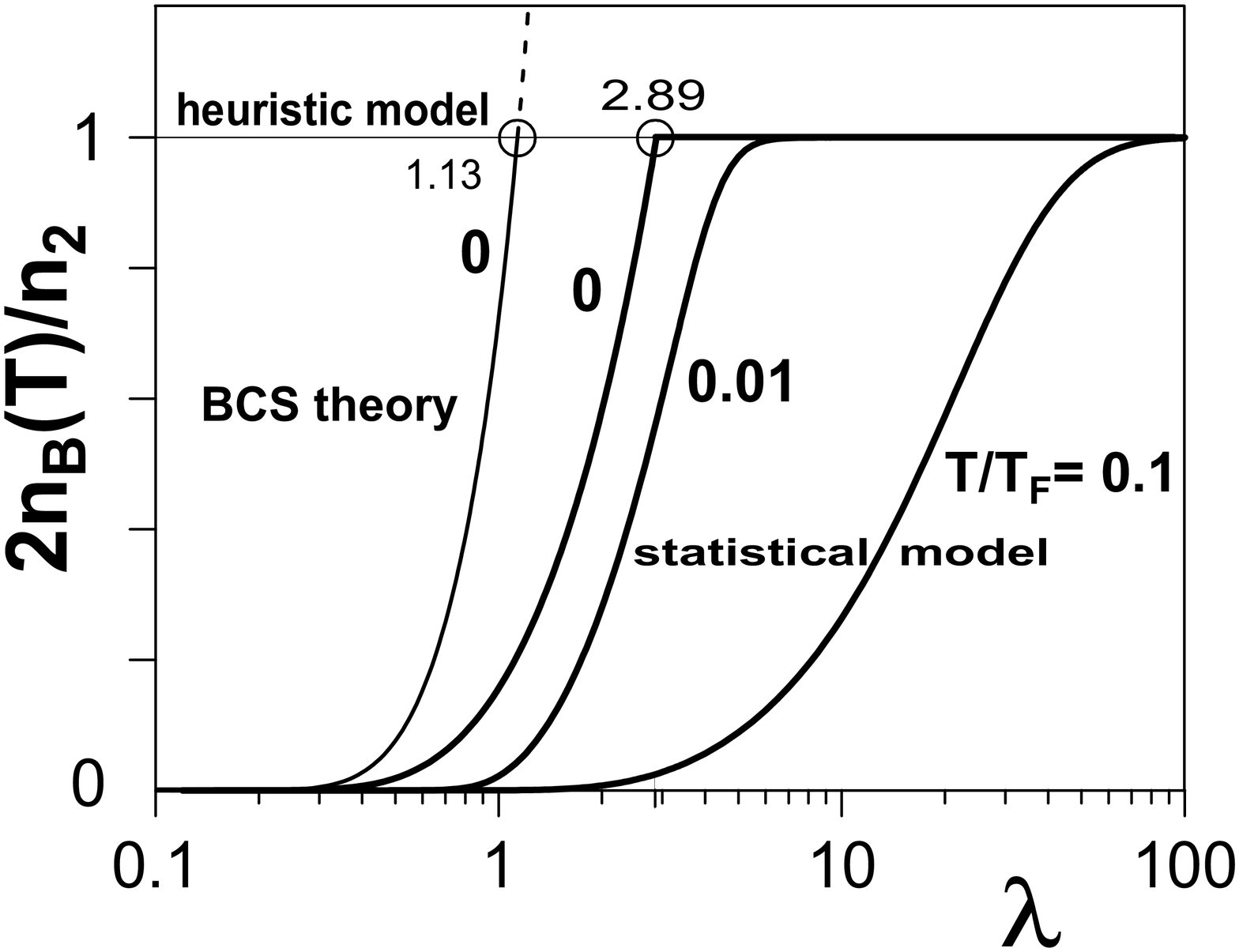,height=2.5in,width=2.7in}
\vspace{0.5cm}
\caption{Fractional number of pairable fermions that are actually paired 
{\it vs}. coupling $\protect\lambda $ for the present statistical model at
three different temperatures (thick curves) and estimated for BCS theory at $T=0$ as explained below (\ref{NB/N2}) (thin curve). The number of pairable
fermions with the BCS model interaction used is just (\ref{n214}); {\it all}
of them are paired at $T=0$ (unrealistically) in the heuristic BEC model,
Ref. {\protect\cite{cas}} Eq. (23). }
\end{center}
\end{minipage}\hfill
\begin{minipage}[b]{2.8in} 
\begin{center}
\psfig{file=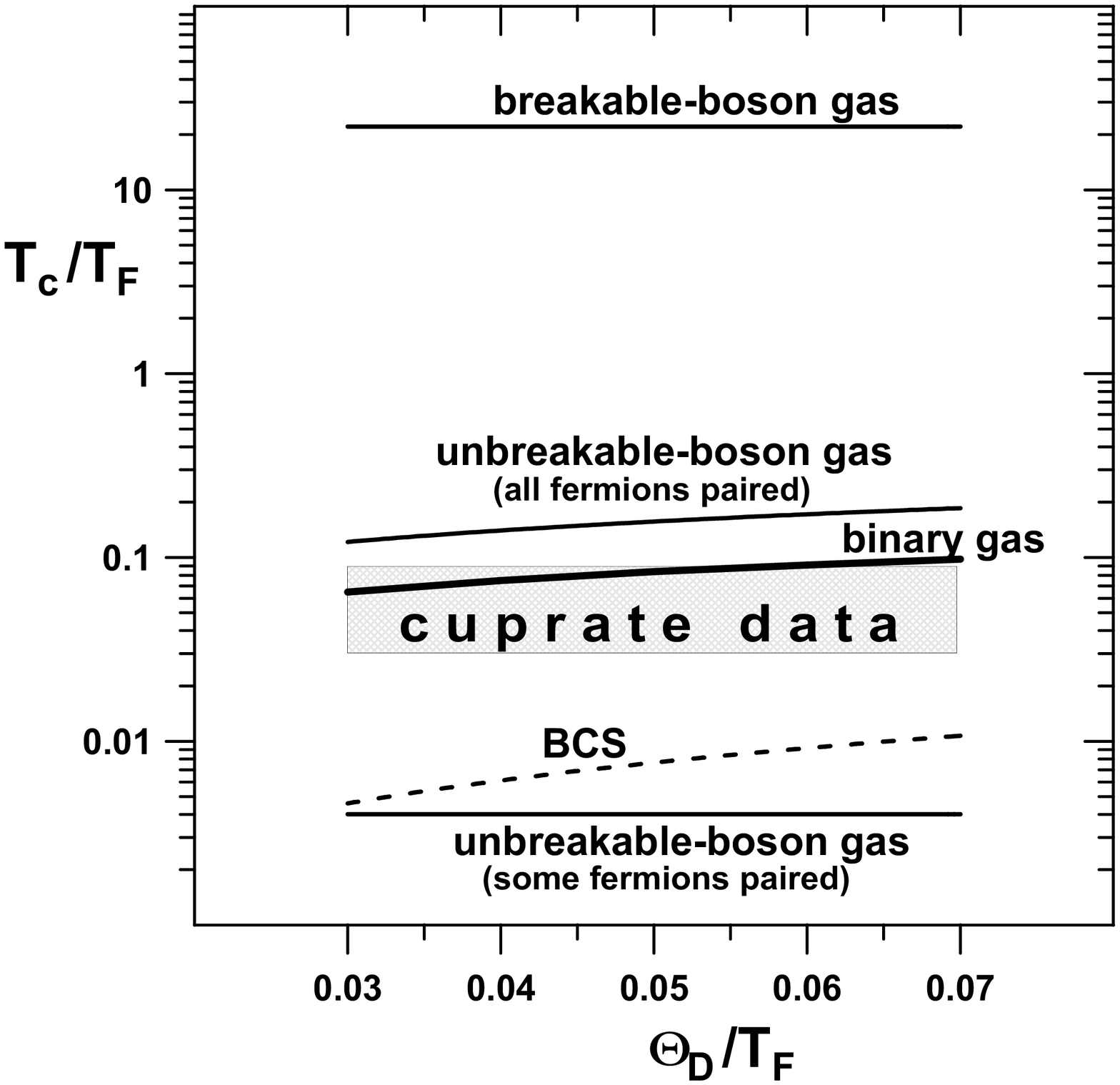,height=3.0in,width=2.7in}
\vspace{-1.5cm}
\caption{ Critical BEC temperature $T_{c}$ in units of $T_{F}$ for the BCS
model interaction with $\protect\lambda =1/2$ for varying $\protect\nu 
\equiv \hbar \protect\omega _{D}/\protect\mu (T_{c})\simeq \Theta _{D}/T_{F}$
for: the pure unbreakable-boson gas with {\it some} and with {\it all}
fermions paired, the former being the solution of (\ref{(11)}) and the
latter taken from Ref. {\protect\cite{cas}}, Eq. (17); for the
breakable-boson gas, from Ref. {\protect\cite{cas}}, Eq. (18); and for the
boson-fermion mixture from (\ref{Tc/nu}) (thick full curve labeled ``binary
gas''). Dashed curve is the BCS theory $T_{c}$, and cuprate experimental
data are taken from Ref. {\protect\cite{Poole}}. }
\end{center}
\end{minipage} 
\label{fig:tau2} 
\end{figure}

If the background unpaired fermions are neglected one has a {\it pure boson
gas} of CPs but with $T$-dependent number density $n_{B}(T)$. Converting the
explicit $T_{c}$-formula (\ref{Tc1}) for $s=1$ and $d=2$\ into an {\it %
implicit} one by allowing $n_{B}$ to be $T$-dependent leaves 
\begin{equation}
T_{c}=\frac{4\sqrt{3}}{\pi ^{3/2}}\frac{\hbar v_{F}}{k_{B}}\sqrt{n_{B}(T_{c})%
},  \label{(11)}
\end{equation}
since $g_{2}(1)\equiv \zeta (2)=\pi ^{2}/6$. \ This requires $n_{B}(T)\equiv
N_{B}(T)/L^{2}$ which from (\ref{N2}) requires (\ref{n20t26}), along with $%
\Delta _{0}(T)$ from (\ref{e13}). Solving these {\it three} coupled
equations simultaneously for $\lambda =1/2$ gives the remarkably constant
value $T_{c}/T_{F}\simeq 0.004$ over the entire range of $\nu \equiv \hbar
\omega _{D}/E_{F}$ values $0.03-0.07$ typical of cuprate superconductors. On
the other hand, the BCS theory formula $T_{c}^{BCS}\simeq 1.13\Theta
_{D}e^{-1/\lambda }$ with $\lambda =1/2$ yields $T_{c}/T_{F}$ $=0.005$ $-$ $%
0.011$ over the same range of $\nu $ values. Unfortunately, both sets of
predictions are well below empirical cuprate values of $T_{c}/T_{F}$ varying 
\cite{Poole} from $0.03-0.09$. \ Pure gas model results \cite{cas}\ for
either breakable or unbreakable bosons {\it without} unpaired fermions are
seen in the figure to overestimate empirical $T_{c}/T_{F}$ values by factors
ranging from two to more than two orders of magnitude. All these results are
wide off the mark.

To obtain the critical temperature {\it without} neglecting the background
unpaired fermions, one needs the exact CP excitation energy dispersion
relation $\varepsilon _{K}(T)\equiv \Delta _{0}(T)-\Delta _{K}(T)$ which is
neither precisely linear in $K$ nor independent of $T$. To determine $\Delta
_{K}(T)$ we need a working equation that generalizes Ref. \cite{PhysC} for $%
T>0$ via the new CP eigenvalue equation (\ref{e13}). For the critical
temperature from the finite-temperature dispersion relation, besides solving
for $\Delta _{K}(T)$, one requires (\ref{n214}), (\ref{N2}) and (\ref{n20t26}%
). \ At $T=T_{c}$ both $N_{B,0}(T_{c})\simeq 0$ and $\mu _{B}(T_{c})\simeq 0$
so that (\ref{N2}) leads \cite{PhysA} to the implicit $T_{c}$-equation for
the {\it binary gas} 
\begin{equation}
1={\frac{\tilde{T}_{c}}{\nu }}\ln \left[ {\frac{1+e^{-\{\tilde{\Delta}_{0}(%
\tilde{T}_{c})/2-\nu \}/\tilde{T}_{c}}}{1+e^{-\{\tilde{\Delta}_{0}(\tilde{T}%
_{c})/2+\nu \}/\tilde{T}_{c}}}}\right] +{\frac{8(1+\nu )}{\nu }}%
\int_{0}^{\kappa _{0}(\tilde{T}_{c})}{\frac{\kappa d\kappa }{e^{[\tilde{%
\Delta}_{0}(\tilde{T}_{c})-\tilde{\Delta}_{\kappa }(\tilde{T}_{c})]/\tilde{T}%
_{c}}-1},}  \label{Tc/nu}
\end{equation}
where quantities with tildes are in units of $\mu (T_{c})\simeq E_{F}$ or $%
T_{F}$, while $\kappa \equiv K/2\sqrt{k_{F}^{2}+k_{D}^{2}}$ and $\nu \equiv
\Theta _{D}/T_{F}$.\ \ {\it Four} coupled equations must now be solved
self-consistently for the exact $T_{c}$ for each $\lambda $ and $\nu $,
including (\ref{e13}) for $\tilde{\Delta}_{0}(\tilde{T})$, and Eq. (35) of
Ref. \cite{PhysA} for both $\tilde{\Delta}_{\kappa }(\tilde{T})$ and $\kappa
_{0}(\tilde{T}_{c})$. Results with $\lambda =1/2$\ labeled ``binary gas'' in
Fig. 2 show a huge enhancement of $T_{c}$, with respect to the
self-consistent result from (\ref{(11)}), arising from the equilibrating
presence of the unpaired fermions and in spite of the very small number of
bosons suggested by Fig. 1 for $\lambda =1/2$. 



For cuprates $d\simeq 2.03$ has been suggested \cite{wen} as more realistic
since it reflects inter-CuO-layer couplings, but our results in that case
would be very close to those for $d=2$ since, e.g., from (\ref{Tc1}) $T_{c}$
for $s=1$ (but {\it not }for $s=2$)\ varies little\ \cite{Sevilla}\ with $d$
around $d=2$. Indeed, if $m_{B\perp }$ and $m_{B}$\ are\ the boson masses 
{\it perpendicular} and {\it parallel}, respectively,{\it \ }to the cuprate
planes, an ``anisotropy ratio'' $m_{B}/m_{B\perp }$ varied from 0 to 1
allows ``tuning'' $d$ continuously from 2 to 3.

Other boson-fermion models \cite{Lee1,Lee2,Rann3,Janko,Rann85,Gesh} have
been introduced, some even addressing \cite{Rann3,Gesh}\ $d$-wave
interaction effects as opposed to the pure $s$-wave considered here, and
some also focusing\ \cite{Rann3,Janko}\ on the pseudogap. \ But calculating
cuprate $T_{c}$ values in quasi-2D without adjustable parameters is not
reported---and indeed predict $T_{c}\equiv 0$ in exactly 2D.

To conclude, a statistical model treating ordinary CPs as non-interacting
bosons in thermal and chemical equilibrium with unpaired fermions yields a
boson number that rises very slowly from zero with coupling, and that
decreases with temperature. When the CP dispersion relation is approximately
linear, it exhibits a BEC of zero-CMM pairs at precisely 2D. Transition
temperatures for the boson-fermion mixture based on the exact CP dispersion
relation for the BCS model electron-phonon interaction are greatly enhanced
over both BCS theory as well as pure-Bose-gas BEC $T_{c}$'s, and are in
rough agreement with empirical cuprate superconductor $T_{c}$'s without any
adjustable parameters. \newline

MC is grateful for partial support from grant PB98-0124, and MdeLl from
PB92-1083 and SAB95-0312, all from DGICYT (Spain), as well as IN102198-9
from PAPIIT (Mexico) and 27828-E from CONACyT (Mexico). MdeLl thanks V.C.
Aguilera-Navarro, P.W. Anderson, D.M. Eagles, R. Escudero, M. Fortes, S.
Fujita, K. Levin, O. Rojo and A.A. Valladares for discussions, V.V.
Tolmachev for correspondence as well, and A. Salazar for help with Fig. 1.

\end{document}